\documentclass[letterpaper]{article} % DO NOT CHANGE THIS
\usepackage{aaai23}  % DO NOT CHANGE THIS
\usepackage{times}  % DO NOT CHANGE THIS
\usepackage{helvet}  % DO NOT CHANGE THIS
\usepackage{courier}  % DO NOT CHANGE THIS
\usepackage[hyphens]{url}  % DO NOT CHANGE THIS
\usepackage{graphicx} % DO NOT CHANGE THIS
\usepackage{natbib}  % DO NOT CHANGE THIS AND DO NOT ADD ANY OPTIONS TO IT
\usepackage{caption} % DO NOT CHANGE THIS AND DO NOT ADD ANY OPTIONS TO IT
\usepackage{cite}
\usepackage{mathrsfs}
\usepackage{multirow}
\usepackage{tabularx}
\usepackage{makecell}
\usepackage{threeparttable}
\usepackage{booktabs}
\usepackage{amsmath}
\usepackage{algorithm}
\usepackage{algorithmic}
\usepackage{color}
\usepackage{etoolbox}
\usepackage{newfloat}
\usepackage{listings}
\DeclareCaptionStyle{ruled}{labelfont=normalfont,labelsep=colon,strut=off} % DO NOT CHANGE THIS
\floatstyle{ruled}
\newfloat{listing}{tb}{lst}{}
\floatname{listing}{Listing}
\title{Self-Supervised Interest Transfer Network via Prototypical Contrastive \\Learning for Recommendation}
\author{
    Guoqiang Sun\textsuperscript{\rm 1,\rm 2},
    Yibin Shen\textsuperscript{\rm 2},
    Sijin Zhou\textsuperscript{\rm 2},
    Xiang Chen\textsuperscript{\rm 1},
    Hongyan Liu\textsuperscript{\rm 1},
    Chunming Wu\textsuperscript{\rm 1},\\
    Chenyi Lei\textsuperscript{\rm 2},
    Xianhui Wei\textsuperscript{\rm 2},
    Fei Fang\textsuperscript{\rm 2}
}
\affiliations{
    \textsuperscript{\rm 1}College of Computer Science and Technology, Zhejiang University, China\\
    \textsuperscript{\rm 2}Alibaba Group, China\\
    \{sungq, wasdnsxchen, hyliu20, wuchunming\}@zju.edu.cn, \\\{shilou.syb, zhousijin.zsj, chenyi.lcy, xianhui.wxh, mingyi.ff\}@alibaba-inc.com
}

\usepackage{bibentry}
\begin{document}

\maketitle

\begin{abstract}
Cross-domain recommendation has attracted increasing attention from industry and academia recently. However, most existing methods do not exploit the interest invariance between domains, which would yield sub-optimal solutions. In this paper, we propose a cross-domain recommendation method: \textbf{S}elf-supervised \textbf{I}nterest \textbf{T}ransfer \textbf{N}etwork (SITN), which can effectively transfer invariant knowledge between domains via prototypical contrastive learning. Specifically, we perform two levels of cross-domain contrastive learning: 1) instance-to-instance contrastive learning, 2) instance-to-cluster contrastive learning. Not only that, we also take into account users' multi-granularity and multi-view interests. With this paradigm, SITN can explicitly learn the invariant knowledge of interest clusters between domains and accurately capture users' intents and preferences. We conducted extensive experiments on a public dataset and a large-scale industrial dataset collected from one of the world's leading e-commerce corporations. The experimental results indicate that SITN achieves significant improvements over state-of-the-art recommendation methods. Additionally, SITN has been deployed on a micro-video recommendation platform, and the online A/B testing results further demonstrate its practical value. Supplement is available at: https://github.com/fanqieCoffee/SITN-Supplement.
\end{abstract}

\section{Introduction}

In single-domain recommender systems, only historical interaction data in one domain are utilized to predict users' behaviors. Earlier work like FM \cite{rendle2010factorization} considers both first-order and second-order feature interactions to boost performance. Some follow-up works such as DIN \cite{zhou2018deep} and DIEN \cite{zhou2019deep} model user behavior sequences with the attention mechanism.\par 
However, there are some critical and non-trivial issues that the single-domain recommendation methods are difficult to address, such as cold-start and data sparsity. Therefore, cross-domain recommendation (CDR) methods emerged and tried to solve these issues. Specifically, these methods aim to achieve higher performance by efficiently utilizing and transferring knowledge among multiple domains. Many CDR methods are dedicated to establishing connections between domains. For instance, CoNet \cite{hu2018conet} establishes cross-connections between base networks to transfer knowledge across domains. With the rapid growth of contrastive learning in computer vision, some methods combine it with CDR. For example, SEMI \cite{lei2021semi} maximizes a lower bound of the mutual information between different domains. Nevertheless, most of existing methods do not explore and exploit the interest invariance between domains. MiNet \cite{ouyang2020minet} just use the attention mechanism to associate the interest of each domain. SEMI only focuses on the transfer of users' individual behaviors. They both ignore the significance of the interest invariance between domains. We argue that there exists a commonality in user interests between domains, which is beneficial for CDR.\par 
\begin{figure*}[ht]
    \centering\includegraphics[scale=0.69]{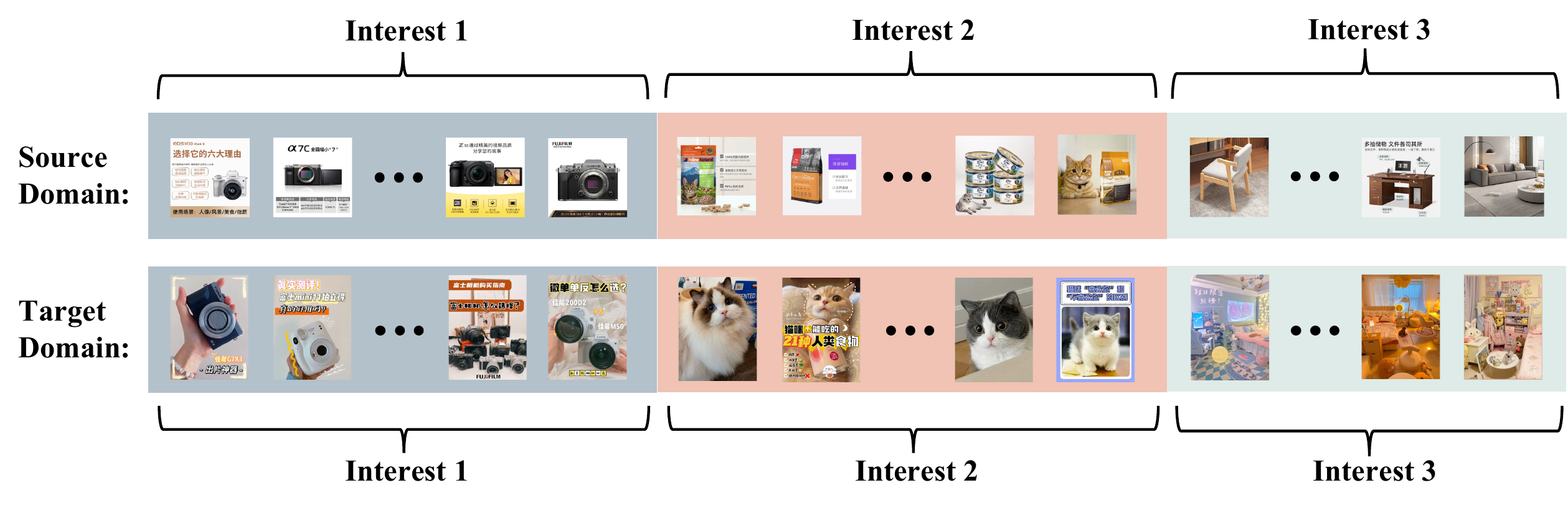}
    \caption{The user clicking sequences in a real-life scenario. The upper part is the user's clicking sequence in a product recommendation platform (source domain), and the lower part is the user's clicking sequence in a content/micro-video recommendation platform (target domain). The user has three interests in both domains and these interests correspond to each other.}
    \label{fig:intro}
\end{figure*}
For instance, Figure \ref{fig:intro} shows the clicking sequences of a user from the source domain and the target domain of a real-life scenario. Here, the source domain is a e-commerce platform where users can purchase products while the target domain is a community where users can post their feelings about the products they purchased and share their lives. In the source domain, we can notice that the user has three interests: 1) cameras, 2) pet supplies (e.g. cat food), 3) home furnishing supplies (e.g. chairs, tables and sofas). Also, in the target domain, we can observe that the user has three similar interests corresponding to the above interests: 1) buying and using guides for cameras, 2) knowledge of pet raising, 3) home design and decoration. We can find that the user has the interest invariance between domains. Yet, existing methods do not explicitly learn such invariant knowledge and transfer it between domains. We believe that model performance will achieve significant improvement if the role of the interest invariance in cross-domain recommendation is fully exploited.\par
Nonetheless, exploiting such knowledge is non-trivial, which raises the following challenges: 1) how to extract the knowledge of interests from each domain; 2) how to transfer the interest invariance between domains; 3) how to make the behaviors and interests of the user work jointly on the model. To address such challenges, we propose \textbf{S}elf-supervised \textbf{I}nterest \textbf{T}ransfer \textbf{N}etwork (SITN), which can efficiently learn and transfer the invariance of users' interests from different domains and allow users' behaviors to fully interact with the interests to obtain higher model performance. Specifically, we treat users' sequential behavioral representations as instances and the interests encoded by the instances as clusters, which can be trained within batches in an end-to-end manner. Next, we design the instance-to-cluster contrastive learning module, which perform prototypical contrastive learning \cite{li2020prototypical} across the source domain and the target domain. It treats an instance and the corresponding cluster from the other domain as a positive pair, to efficiently learn and transfer the interest invariance between domains. Meanwhile, we perform contrastive learning on the representations of users' behaviors from different domains like SEMI and we called it instance-to-instance contrastive learning. By employing the two levels of contrastive learning simultaneously, the instance-to-instance and instance-to-cluster contrastive learning can fully interact with each other and work jointly on the model, therefore, improve the model performance. In addition to that, a user's interests are often diverse and can not be determined by a single interest alone. Therefore, we introduce the multi-view interest module which reflects the user's varied interests based on the instance-to-cluster contrastive learning. Last but not least, there are numerous distinct levels of granularity for a user's interests across different interest spaces. For instance, the granularity of a girl's fashion interest space will typically be much finer than that of the digital interest space. In order to remedy the problem, we propose the multi-granularity module, which concentrates on users' different granular interests. The offline experimental results suggest the effectiveness of SITN, and the online A/B testing results further highlight its usefulness.\par
In summary, we can conclude our main contributions as follows:
\begin{itemize}
    \item We underline the significance of the interest invariance across domains and explore how it plays a role in cross-domain recommendation.

    \item We propose the model SITN, which efficiently learn and transfer the interest invariance between domains, and also take into account users' multi-view and multi-granularity interests.

    \item We conduct extensive offline experiments on a large-scale industrial dataset and a public dataset. The results show significant improvements over state-of-the-art methods.
    
    \item We have deployed SITN on a micro-video recommendation platform and conducted online A/B tests to show its practical value. More details are in the supplement.
\end{itemize}

\section{Related Work}
\subsection{Cross-domain Recommendation}
The basic idea of cross-domain recommendation (CDR) is to introduce knowledge from other domains to alleviate the problems such as cold-start and data sparsity, not only that, also to strengthen user preferences. Therefore, the key to CDR is how to leverage knowledge from domains effectively and establish connections between them efficiently. In the early days, CDR was mainly based on some traditional methods. The collective matrix factorization model \cite{singh2008relational} factorizes rating matrices collectively of two domains into user representations and item representations. Besides, the factorization machines-based model named FM-MCMC \cite{loni2014cross} takes into account the user's behaviors in the source domain and target domain simultaneously during the training process. Over the last few years, with the rapid growth of deep learning, some methods based on it have sprung up. The collaborative cross-network (CoNet)  \cite{hu2018conet} establishes cross-connections between base networks to enable dual knowledge transfer. Compared with CoNet, DDTCDR \cite{li2020ddtcdr} learns a latent orthogonal mapping function between domains, which not only preserves the similarity of users' preferences, but also effectively calculates the reverse mapping function. There is also a method called Mixed Interest Network (MiNet) \cite{ouyang2020minet} that transfers users' interests from source domain to target domain by using attention mechanisms. To remedy the cold-start problem, PTUPCDR \cite{zhu2022personalized} utilizes a meta network to transfer the user preferences between domains. The method most similar to ours is SEMI \cite{lei2021semi}, which applies contrastive learning to CDR, bridging the gap between domains by pulling the representations of user behavior sequences in different domains closer. However, SEMI only performs contrastive learning at the instance-to-instance level, neglecting their interests, which will introduce noise and lead to a sub-optimal solution. To this end, our method employs contrastive learning at both instance-to-instance and instance-to-cluster levels simultaneously, which could better leverage and transfer the knowledge between domains. 
\subsection{Contrastive Learning}
Contrastive learning (CL) plays a significant role in the self-supervised learning paradigms, aiming to extract essential invariant knowledge by constructing the pretext tasks through the principle of mutual information maximization \cite{oord2018representation,velickovic2019deep}, which achieved great success in computer vision \cite{chen2020simple,he2020momentum}, then extended to natural language processing \cite{gao2021simcse} and graph representation learning \cite{sun2019infograph,you2020graph,zhu2021graph}. Recently, more and more methods related to CL have emerged in the field of recommender systems (RS) \cite{zhou2020s3,wu2021self,xie2022contrastive}. Specifically, the main idea of contrastive learning is to construct pairs of positive and negative samples by augmenting the data, then pull the former closer and push the latter farther in the feature space. At early stages, a variety of works focused on constructing different pretext tasks through different ways of augmentation. For instance, one is to exploit feature correlations and augment data by feature masking \cite{yao2021self}. Additionally, a method known as prototypical contrastive learning (PCL) \cite{li2020prototypical}, which was recently applied to RS, treats an instance and the cluster to which it belongs as a positive pair.
% There is also a method that treats instance and the cluster to which it belongs as positive pair, which is called prototypical contrastive learning (PCL) \cite{li2020prototypical} and was applied to RS recently.
One of the PCL-based methods in RS is to construct positive sample pairs by semantic neighbors referring to semantically similar neighbors which may not be directly reachable on graphs in graph collaborative filtering \cite{lin2022improving}. The other method \cite{chen2022intent} in Sequential Recommendation, which is similar to our work, treats the clusters of behavior sequences as users' intents like prototypes.
% The other method in Sequential Recommendation is to treat the clusters of behavior sequences as users' intents like prototypes which is similar to our work. 
While their method belongs to single-domain recommendation, ours is used for cross-domain recommendation, and the interest clusters are trainable within batches. Not only that, we also consider the user's multi-granularity and multi-view interests. 

\section{Preliminaries and Notations}
We have a set of users and items which can be denoted by $\mathcal{U}$ and $\mathcal{V}$ respectively. Let $\mathcal{S}^u = \left\{s^u_1, s^u_2,\cdots, s^u_{|\mathcal{S}^{u}|}\right\}$ and $\mathcal{T}^u = \left\{t^u_1, t^u_2, \cdots , t^u_{|\mathcal{T}^{u}|}\right\}$ be clicking sequences sorted in chronological order from source domain and target domain respectively for each user $u \in \mathcal{U}$, where $|\mathcal{S}^{u}|$ and $|\mathcal{T}^{u}|$ denote the length of each sequence. Here, $s^u_i$ and $t^u_j$ are the i-th and j-th items clicked by the user $u$ in the source and target domain respectively. Additionally, we define interest cluster matrix $\begin{matrix} C_{K\times D}^s \end{matrix}$ and $\begin{matrix} C_{K\times D}^t \end{matrix}$ for source domain and target domain respectively, where $K$ and $D$ denote the number of clusters and the dimension of each cluster respectively which are hyper-parameters. Note that $\begin{matrix} C_{K\times D}^s \end{matrix}$ and $\begin{matrix} C_{K\times D}^t \end{matrix}$ are randomly initialized and trainable. The task in our scenario is to predict the probability of user $u \in \mathcal{U}$ potentially clicking the item $v \in \mathcal{V}$ from target domain given sequences $\mathcal{S}^u$ and $\mathcal{T}^u$.

\section{Method}
\begin{figure*}[ht]
    \centering\includegraphics[scale=0.63]{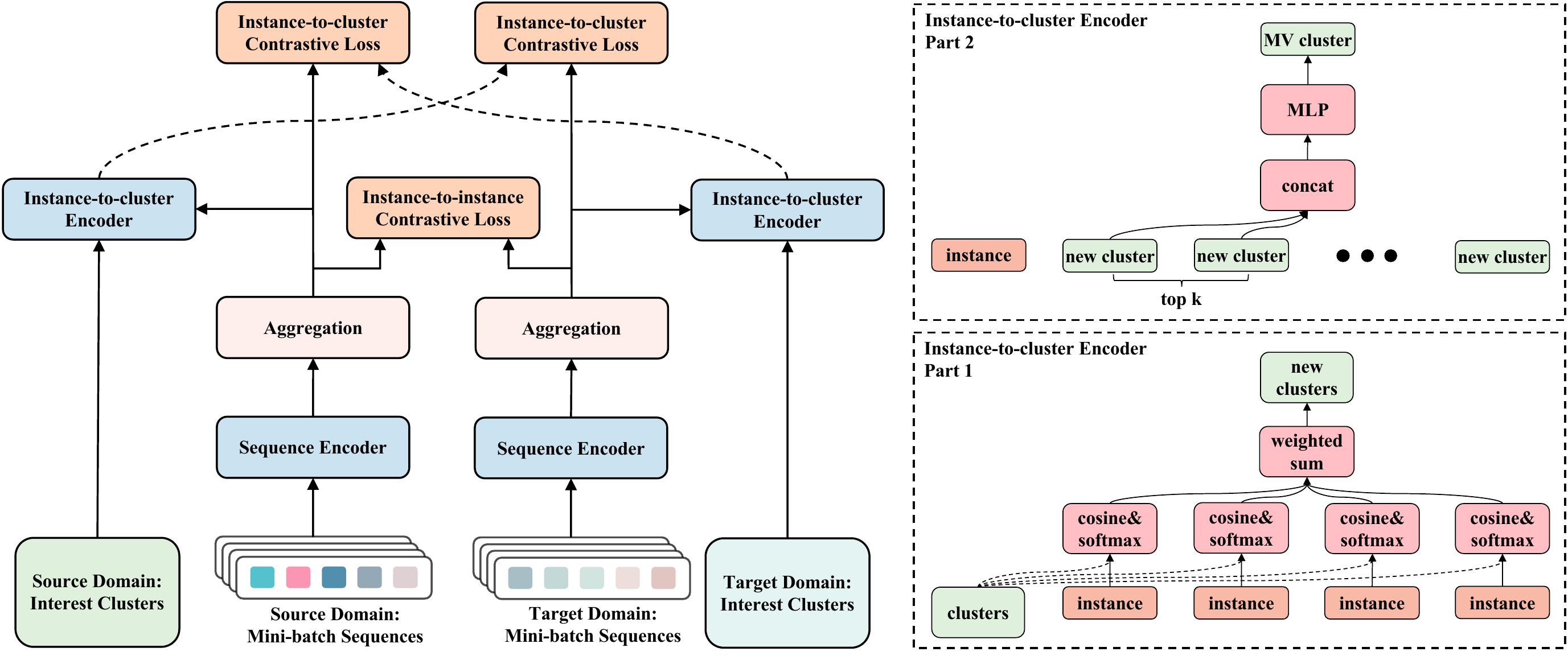}
    \caption{The overall framework of SITN's contrastive learning module. The left part shows the contrastive learning module at a high level. The right part shows the instance-to-cluster encoder. The role of the first part of instance-to-cluster encoder is to generate new clusters from instances and original clusters. The lower right part shows a example of clusters and four instances. The role of the second part of instance-to-cluster encoder is to generate MV clusters from instances and new clusters. The upper right part shows a example of new clusters and one instance, and the value of k is two.}
    \label{fig:structure}
\end{figure*}

In this section, we first introduce the significant basic modules of \textbf{S}elf-supervised \textbf{I}nterest \textbf{T}ransfer \textbf{N}etwork (SITN) (shown in Figure \ref{fig:structure}) , which incorporated three main modules: 1) \emph{Sequence encoder and aggregation layer}, which captures users' sequential behavior patterns and obtains the representations of users' encoded clicking sequences (instances); 2)  \emph{Instance-to-instance contrastive learning}, which performs contrastive learning on the instances from different domains; 3) \emph{Instance-to-cluster contrastive learning}, which performs prototypical contrastive learning on the instances and the interest clusters from different domains. Next, we elaborate our two-stage training paradigm. In the first stage, we employ InfoNCE \cite{oord2018representation} in a self-supervised learning manner. In the second stage, instead of training from scratch, we use and fine-tune the pre-trained sequence encoders from the previous stage to predict the probability of a user would click a specific item in the target domain.
\subsection{Sequence Encoder and Aggregation Layer}
We employ multi-head attention mechanism \cite{vaswani2017attention}, analogous to many other approaches \cite{chen2019behavior,de2021transformers4rec}, to encode sequences in the source domain and the target domain, enabling the model to concentrate on critical information and thoroughly learn to assimilate it.
In this work, we get two clicking sequences from source domain and target domain for each user. 
Here, we use the self-attention mechanism, which means that Query, Key and Value are all the same (the clicking sequence). Then, we perform scaled dot-product attention and compute the matrix of outputs as follows:
\begin{equation}
Attention(Q,K,V) = softmax(\frac{Q^TK}{\sqrt{d_k}})V,
\label{eq:attention}
\end{equation}
where $Q \in \mathcal{R}^d, K \in \mathcal{R}^d$ and $V \in \mathcal{R}^d$ represent Query, Key and Value  respectively, in addition, $d_k$ is the dimension of $K$. Multi-head attention mechanism is beneficial to learn relevant information in different representation subspaces. Therefore we linearly project Query, Key and Value h times with different trainable weight matrices and then perform attention function on them to compute outputs. Next, the all outputs are concatenated and projected again to get the final outputs.
After the sequence encoder layer, we perform aggregation layer (e.g., mean pooling function) to get the representations of encoded sequences.
\subsection{Instance-to-instance Contrastive Learning}
In this module, We perform instance-to-instance contrastive learning (CL-I2I), similar to SEMI \cite{lei2021semi}, on the representations of encoded clicking sequences between domains.
By pulling the representations of encoded clicking sequences of the same user closer in different domains, the goal of CL-I2I is to extract invariant behavior patterns and preferences.
Specifically, as shown in Figure \ref{fig:structure}, let $\mathcal{P}^s = \left\{p^s_1, p^s_2,\cdots, p^s_n\right\}$ and $\mathcal{P}^t = \left\{p^t_1, p^t_2,\cdots, p^t_n\right\}$ denote the representations of encoded sequences in the source domain and target domain respectively within a mini-batch, where $n$ is the size of mini-batch. For same user, the subscripts of $p^s$ and $p^t$ will be the same. Therefore, we could obtain the outputs (the representations of encoded sequences) as follows:
\begin{equation}
    p^s_i = g(f_s(x^s_i)),\qquad       
    p^t_i = g(f_t(x^t_i)), 
\label{eq:encoder}
\end{equation}
where $f^s(\cdot)$ and $f^t(\cdot)$ are the encoders and $g(\cdot)$ is the mean pooling function described in the previous subsection, $x^s_i$ and $x^t_i$ are the identical user's clicking sequences in the source domain and target domain respectively as inputs of the encoders. Next, we perform CL-I2I with infoNCE loss,
\begin{equation}
    \mathcal{L}_{i2i}^{s2t} = -\sum_{k=1}^nlog\frac{exp(sim(p^s_k,p^t_k)/\tau)}{\sum_{i=1}^nexp(sim(p^s_k,p^t_i)/\tau)},
\label{eq:i2i_s2t_loss}
\end{equation}
where $sim(\cdot)$ is a dot product and $\tau$ denotes the the hyper-parameter, known as \emph{temperature} in softmax. CL loss is supposed to be symmetrical, so analogously, we could compute $\mathcal{L}_{i2i}^{t2s}$ as follows:
\begin{equation}
    \mathcal{L}_{i2i}^{t2s} = -\sum_{k=1}^nlog\frac{exp(sim(p^t_k,p^s_k)/\tau)}{\sum_{i=1}^nexp(sim(p^t_k,p^s_i)/\tau)}.
\label{eq:i2i_t2s_loss}
\end{equation}
Combining the two losses, we could obtain instance-to-instance contrastive loss as $\mathcal{L}_{i2i} = \mathcal{L}_{i2i}^{s2t} + \mathcal{L}_{i2i}^{t2s}$.
\subsection{Instance-to-cluster Contrastive Learning}
Considering solely CL-I2I without the user's interests adds noise and yields sub-optimal solution.  
Motivated by such issue, we propose instance-to-cluster contrastive learning (CL-I2C), which aims to transfer knowledge more effectively and boost performance by exploiting the same user's interest invariance across different domains.
Furthermore, CL-I2I and CL-I2C could interact with each other to jointly enhance robustness of the model and achieve higher performance. In this module, we first introduce the instance-to-cluster encoder, which could obtain multi-view clusters from instances. Then we elaborate the instance-to-cluster contrastive loss, which is the key component in our method.
\subsubsection{Instance-to-cluster Encoder}
We define and randomly initialize the interest cluster matrix $\begin{matrix} {C}_{K\times{D}}^s \end{matrix}$ and $\begin{matrix} {C}_{K\times{D}}^t \end{matrix}$ for source domain and target domain where each row represents a interest cluster. Given $\begin{matrix} {C}_{K\times{D}}^s \end{matrix}$, $\begin{matrix} {C}_{K\times{D}}^t \end{matrix}$ and the representations of encoded sequences $\mathcal{P}^s$ and $\mathcal{P}^t$, similar to TCC \cite{shen2021you}, as shown in the lower right part of Figure \ref{fig:structure}, we can compute the new interest clusters in the source domain:
\begin{equation}
\begin{aligned}
u_k^s &= \sum_{i=1}^{n}\pi_i^s(k)\cdot{p^s_i},\\
\pi^s_i(k) &= \frac{exp(sim(c^s_k,p^s_i))}{\sum_{k'=1}^K{exp(sim(c^s_{k'},p^s_i))}},
\end{aligned}
\label{eq:i2c_encoder}
\end{equation}
where $u_k^s$ and $c^s_k$ denote the k-th new and original interest cluster in the source domain respectively. As a result, we could get the new interest cluster matrix $\begin{matrix} {U}_{K\times{D}}^s \end{matrix}$ in the source domain. Meanwhile, the new interest cluster matrix $\begin{matrix} {U}_{K\times{D}}^t \end{matrix}$ in the target domain could be obtained by the same way. We further consider the diversity of user interests, thus we build positive sample pairs with the k interests that are most related to the user’s behaviors to generate soft labels, a method we refer to as multi-view CL-I2C. In particular, as shown in the upper right part of Figure \ref{fig:structure}, we obtain the multi-view clusters (MV clusters) $\mathcal{Q}^s = \left\{q_1^s,q_2^s,\cdots,q_n^s\right\}$ and $\mathcal{Q}^t = \left\{q_1^t,q_2^t,\cdots,q_n^t\right\}$ which represent the users' multi interests from the source and target domain respectively:
\begin{equation}
\begin{aligned}
q_i^s &= MLP(Concat(u^s_{i1},u^s_{i2},\cdots,u^s_{ik})),\\
q_i^t &= MLP(Concat(u^t_{i1},u^t_{i2},\cdots,u^t_{ik})),
\end{aligned}
\label{eq:multi_interest}
\end{equation}
where $\left\{u^s_{i1},u^s_{i2},\cdots,u^s_{ik}\right\}$ and $\left\{u^t_{i1},u^t_{i2},\cdots,u^t_{ik}\right\}$ denote the top k interests which are most similar to the user's encoded clicking sequences in the source domain and target domain respectively. In this manner, we could train the cluster prototypes end-to-end, requiring no alternating steps.
\subsubsection{Instance-to-cluster Contrastive Loss}
Since users' behavior patterns are consistent with their interests, it is necessary to perform contrastive learning on the encoded clicking sequences and interest clusters. In our method, for the purpose of fully exploiting the interest invariance and transferring the knowledge efficiently between domains, we perform CL-I2C across domains. Each instance forms a positive pair with its corresponding MV cluster, and negative pairs with all new clusters. Here, the instance and the clusters belong to different domains. Therefore, comparable to instance-to-instance contrastive loss, we can compute instance-to-cluster contrastive loss in one interest subspace as:
\begin{equation}
\begin{small}
\begin{aligned}
    e_{i2c}^{s2t} &= -\sum_{k=1}^nlog\frac{exp(sim(p^s_k,q^t_k)/\tau)}{exp(sim(p^s_k,q^t_k)/\tau)+\sum_{i=1}^Kexp(sim(p^s_k,u^t_i)/\tau)},\\
    e_{i2c}^{t2s} &= -\sum_{k=1}^nlog\frac{exp(sim(p^t_k,q^s_k)/\tau)}{exp(sim(p^t_k,q^s_k)/\tau)+\sum_{i=1}^Kexp(sim(p^t_k,u^s_i)/\tau)},\\
    e_{i2c} &= e_{i2c}^{s2t} + e_{i2c}^{t2s}.
\end{aligned}
\end{small}
\label{eq:i2c_loss}
\end{equation}
In addition to that, users’ interests in different subspaces are multi-granular. For instance, a girl may have a finer granularity in the fashion interest space than in the digital interest space. Therefore, in order to take into account users' interests in different subspaces, we also perform multi-granularity CL-I2C. For interests in different subspaces, we can obtain the instance-to-cluster contrastive loss $\mathcal{E}_{i2c} = \left\{e_{i2c}^1,e_{i2c}^2,\cdots,e_{i2c}^l\right\}$ where $l$ is the number of different subspaces which is a hyper-parameter. Finally, we can get the instance-to-cluster contrastive loss as $\mathcal{L}_{i2c} = \sum_{i=1}^{l}e_{i2c}^i$.

\subsection{Two-stage Training Paradigm}
In the first stage, we train the model in a self-supervised learning manner using the contrastive loss $\mathcal{L}_{ssl} = \mathcal{L}_{i2i} + \mathcal{L}_{i2c}$. Our task is to predict whether the user clicks a specific item in the target domain. Thus, our goal is to extract the user's interests based on the user's historical behaviors in the two domains and explore the connections between the interests and the specific item in the target domain. However, the user's interactions in the target domain are often much more sparse than in the source domain. Nonetheless, we can effectively address such problem with the first stage of self-supervised learning of SITN. In the first stage, we learn and transfer the user's interest invariance between domains via contrastive learning, which allows us to accurately capture the user's preferences in the subsequent training. Instead of training from scratch, in the second stage, we initialize sequence encoders with the weights obtained in the previous stage.\par
To be specific, in the second stage, we first obtain the user's behavioral sequence representations $\mathcal{Z}^s = \left\{z^s_1, z^s_2,\cdots, z^s_{|\mathcal{Z}^{s}|}\right\}$ and $\mathcal{Z}^t = \left\{z^t_1, z^t_2,\cdots, z^t_{|\mathcal{Z}^{t}|}\right\}$ encoded by the sequence encoders which are pre-trained in the first stage from the source domain and target domain respectively, where $|\mathcal{Z}^s|$ and $|\mathcal{Z}^t|$ denote the length of each sequence. Then, we can obtain the user representation $u_f^s$ in the source domain by feeding the target item $v$ and $\mathcal{Z}^s$ into a target attention block to capture the connection between the target item and user behavior sequence, where Query is $v$, and key and value are the same, that is $\mathcal{Z}^s$. Similarly, we can get the user representation $u_f^t$ in the target domain. Next, we concatenate $u_f^s$, $u_f^t$ and $v$, and feed them into a MLP block to predict the probability $\hat{y}$ whether the user would like to click the item in the target domain as follows:
\begin{equation}
\begin{split}
\begin{aligned}
  \hat{y} &= MLP(Concat(u_f^s, u_f^t, v)),
\end{aligned}
\end{split}
\label{eq:logits}
\end{equation}
where MLP denotes multi-layer perceptron. Finally, the cross-entropy loss can be computed as:
\begin{equation}
\begin{split}
\begin{aligned}
    \mathcal{L}(y,\hat{y}) = -(ylog\hat{y} + (1-y)log(1-\hat{y})),
\end{aligned}
\end{split}
\label{eq:celoss}
\end{equation}
where $y \in \left\{0, 1\right\}$ is the ground-truth label indicating whether the
user clicks the item in the target domain or not.

\section{Experiments}
\subsection{Datasets}
To verify the effectiveness of the proposed method, we conduct extensive experiments on two datasets. The statistics of the datsets are shown in Table \ref{tab:data}.
\subsubsection{Industrial Dataset}
We collect a large-scale dataset from one of the world’s leading e-commerce corporations. In the dataset, the source domain is a e-commerce platform where users can purchase products while the target domain is a community where users can post their feelings about the products they purchased and share their lives. We collect the logs on the corporation from seven consecutive days during May 2022. The dataset contains user ID, item ID, item title, item category and clicking sequences from the source domain and target domain. Additionally, we only reserve those logs whose clicking sequence lengths in the two domains are both larger than 10 and truncate the clicking sequence length to 100.
\subsubsection{Amazon Dataset}
The Amazon review dataset \cite{ni2019justifying} is a widely used public dataset for recommender systems. The dataset contains users' review data and products' metadata from many domains (Books, Movies and TV, Digital music, Electronics, etc.). Here, we take the book domain as the source domain and the movie \& TV domain as the target domain. For each domain, we consider the ratings of 4 or more as clicking behaviors and form the clicking sequences by sorting the logs in chronological order. In addition to the clicking sequences, the features used include user ID, item ID, item title, item brand, item main and secondary category. Moreover, we only keep those logs with book clicking sequence length of 5 or more and truncate the clicking sequence length to 100.

\newcommand{\tabincell}[2]{\begin{tabular}{@{}#1@{}}#2\end{tabular}}
\begin{table}[h]
    \setlength{\tabcolsep}{5pt}
 \centering
 \small
    \begin{tabularx}{0.45\textwidth}{p{3cm}|X|X}
    \toprule
    \makecell[c]{\text{Dataset}}&\makecell[c]{\textbf{Industrial}}&\makecell[c]{\textbf{Amazon}}\cr
    \hline
    \hline
    \makecell[c]{\text{\# Shared Users}}&\makecell[c]{23,706,610}&\makecell[c]{87,896}\cr

    \makecell[c]{\text{\# Items(Source)}}&\makecell[c]{54,829,040}&\makecell[c]{673,826}\cr
    \makecell[c]{\text{\# Items(Target)}}&\makecell[c]{13,149,185}&\makecell[c]{100,164}\cr
    \makecell[c]{\text{\# Instances}}&\makecell[c]{445,821,389}&\makecell[c]{1,290,358}\cr
    \makecell[c]{\text{Avg. \# clicked(Source)}}&\makecell[c]{94}&\makecell[c]{39}\cr
    \makecell[c]{\text{Avg. \# clicked(Target)}}&\makecell[c]{60}&\makecell[c]{36}\cr
    \bottomrule[0.8pt]
    \end{tabularx}
    \caption{Statistics of two utilized datasets. (avg. - average)}
    \label{tab:data}
\end{table}

\begin{table}[!t]
\setlength{\tabcolsep}{3pt}
\renewcommand{\arraystretch}{1.2}
% \vskip -10pt
\label{tab_auc}
\centering
\scalebox{0.92}{
\begin{tabular}{|l|l|c|c||c|c|}
\hline
\multicolumn{2}{|c|}{} & \multicolumn{2}{c||}{\textbf{Industrial}} & \multicolumn{2}{c|}{\textbf{Amazon}} \\
\hline
\multicolumn{2}{|c|}{\textbf{Model}} & AUC & Logloss & AUC & Logloss \\
\hline
\multirow{7}{*}{\tabincell{c}{Single-\\domain}} 
& FM & 0.6501 & 0.4486 &0.6215 & 0.4624 \\
& Wide\&Deep & 0.6939 & 0.3701 & 0.7154 & 0.3411 \\
& DeepFM & 0.7040 & 0.3672 & 0.7201 & 0.3358 \\
& DIN & 0.7173 & 0.3408 & 0.7370 & 0.3183 \\
& DIEN & 0.7185 & 0.3392 & 0.7397 & 0.3062 \\
& CLSR & 0.6935 & 0.3740 & 0.6752 & 0.4061 \\
\hline
\multirow{5}{*}{\tabincell{c}{Cross-\\domain}} 
& MV-DNN & 0.7169 & 0.3391 & 0.7293 & 0.3285 \\
& CoNet & 0.7188 & 0.3382 & 0.7275 & 0.3278 \\
& MiNet & 0.7218 & 0.3313 & 0.7305 & 0.3260 \\
& SEMI & 0.7304 & 0.3276 & 0.7439 & 0.3036 \\
& SITN & \textbf{0.7351}* & \textbf{0.3207}* & \textbf{0.7525}* & \textbf{0.2857}* \\
\hline
\end{tabular}}
\vskip -8pt
\caption{Test AUC and Logloss. * indicates the statistical significance for $p<=0.01$ compared with the best baseline method based on the paired t-test.}
\label{tab:method}
\end{table}

\subsection{Overall Performance}
We choose six single-domain methods and four cross-domain methods to compare with ours as follows:
\subsubsection{Single-Domain Methods}
    \begin{itemize}
    \item \textbf{FM} \cite{rendle2010factorization}: This method considers both first-order and second-order feature interactions.
    \item \textbf{Wide\&Deep} \cite{cheng2016wide}: This method uses both logistic regression and deep neural networks (DNN).
    \item \textbf{DeepFM}\cite{guo2017deepfm}: The method combines FM and DNN.
    \item \textbf{DIN}\cite{zhou2018deep}: This work uses the attention mechanism in sequential recommendation.
    \item \textbf{DIEN} \cite{zhou2019deep}: This work extends DIN with interest extractor layer and interest evolution layer.
    \item \textbf{CLSR} \cite{zheng2022disentangling}: This work models the long and short term sequences of users by using the idea of contrastive learning.
    \end{itemize}
\subsubsection{Cross-Domain Methods}
    \begin{itemize}
    \item \textbf{MV-DNN} \cite{elkahky2015multi}: This work is based on Deep Structured Semantic Model \cite{huang2013learning} and combines the features from multiple domains.
    \item \textbf{CoNet} \cite{hu2018conet}: This work establishes cross-connections between base networks to transfer knowledge across domains. 
    \item
    \textbf{MiNet}\cite{ouyang2020minet}: The model uses the attention mechanism to extract and transfer the knowledge of interests between domains.
    \item \textbf{SEMI}\cite{lei2021semi}: This work employs contrastive learning to encourage the representations of sequences in the source domain and target domain to be similar.
    \end{itemize}
We perform extensive experiments on both datasets and use the Area Under the ROC Curve (AUC) and the value of Eq. \ref{eq:celoss} as the evaluation metrics for the methods. According to Table \ref{tab:method}, the following three conclusions can be drawn: 
\subsubsection{Our proposed method achieves the best performance on both datasets.}
We can notice that SITN outperforms all of the baselines on both datasets. The next best method is SEMI, which is a method similar to ours. SITN exceeds it by 0.64\% and 1.16\% on the industrial dataset and amazon dataset respectively in terms of AUC. The results indicate the effectiveness of SITN, thanks to the fact that it fully learns and efficiently transfers the knowledge of interests between domains.
\subsubsection{The cross-domain methods perform better overall than the single-domain methods on both datasets.}
Based on the experimental results, the cross-domain methods are significantly better than the single-domain methods on the industrial dataset, and are also better overall on the amazon dataset except for DIN and DIEN. We argue that the cross-domain methods can effectively transfer some valid knowledge between domains, especially in real complicated industrial-level scenarios. Among these cross-domain methods, our proposed method is still the most efficient one.
\subsubsection{Contrastive learning is beneficial for cross-domain recommendation.}
We can observe that our method and SEMI give the best results on the two datasets, which are both based on contrastive learning. The experimental results further demonstrate that contrastive learning does play a crucial role in cross-domain recommendation. We assume that is due to the fact that contrastive learning can extract and transfer some valid knowledge between domains, which is useful for capturing users' preferences.
\subsection{Ablation Study}
We conduct two ablation studies to further explore the effect of instance-to-cluster contrastive learning and the effects of multi-granularity and multi-view modules.
\subsubsection{Effect of Instance-to-cluster Contrastive Learning}
\begin{table}[h]
 \setlength{\tabcolsep}{5pt}
 \centering
 \small
  {
    \begin{tabularx}{0.45\textwidth}{p{2cm}lXXlXX}
    \toprule
    \multirow{3}{*}{\text{Method}}&&\multicolumn{2}{c}{\text{Industrial}}&&\multicolumn{2}{c}{\text{Amazon}}\cr
    \cline{3-4}  \cline{6-7}
    &&\text{AUC}&\text{Logloss}&&\text{AUC}&\text{Logloss}\cr
    \hline
    w/o $\mathcal{L}_{i2c}$\&$\mathcal{L}_{i2i}$&&0.7203&0.3325&&0.7328&0.3242\cr
    w/o $\mathcal{L}_{i2c}$&&0.7295&0.3304&&0.7456&0.3020\cr
    w/o $\mathcal{L}_{i2i}$&&0.7324&0.3259&&0.7489&0.2959\cr
    \hline
    SITN&&\textbf{0.7351}	&\textbf{0.3207}&&\textbf{0.7525}&\textbf{0.2857}\cr
    \hline
    \bottomrule
    \end{tabularx}
    }
    \caption{Effect of instance-to-cluster contrastive learning.}
    \label{tab:ab1}
\end{table} 
Motivated by the desire to exploit the interest invariance between domains, we introduce instance-to-cluster contrastive learning. In this module, we investigate whether the design of CL-I2C fully utilizes of its role and jointly enhance robustness of the model with CL-I2I. To be specific, we remove CL-I2C, CL-I2I and both of them from SITN respectively and conduct extensive experiments on the two datasets. As shown in Table \ref{tab:ab1}, we can find that removing either CL-I2I or CL-I2C will result in performance degradation. And if they are both removed, performance degradation is even greater. We can conclude that utilizing solely CL-I2I is somewhat effective in capturing user preferences and enhancing the model's performance, but not as effective as using only CL-I2C. Further, we can obtain the best performance by exploiting both CL-I2I and CL-I2C. We argue that, with the full interaction and co-action of CL-I2I and CL-I2C, which allows the interests between domains to be learned and transferred efficiently, the model will get higher performance.
\subsubsection{Effects of Multi-granularity and Multi-view Modules}
\begin{table}[h]
 \setlength{\tabcolsep}{5pt}
 \centering
 \small
  {
    \begin{tabularx}{0.45\textwidth}{p{2cm}lXXlXX}
    \toprule
    \multirow{3}{*}{\text{Method}}&&\multicolumn{2}{c}{\text{Industrial}}&&\multicolumn{2}{c}{\text{Amazon}}\cr
    \cline{3-4}  \cline{6-7}
    &&\text{AUC}&\text{Logloss}&&\text{AUC}&\text{Logloss}\cr
    \hline
    w/o MG\&MV&&0.7310&0.3343&&0.7462&0.3013\cr
    w/o MG&&0.7335&0.3236&&0.7493&0.2953\cr
    w/o MV &&0.7327&0.3305&&0.7475&0.2965\cr
    \hline
    SITN&&\textbf{0.7351}	&\textbf{0.3207}&&\textbf{0.7525}&\textbf{0.2857}\cr
    \hline
    \bottomrule
    \end{tabularx}
    }
    \caption{Effects of multi-granularity and multi-view modules.}
    \label{tab:ab2}
\end{table}
The user preference is typically closely tied to multiple of the most similar interest clusters rather than being solely defined by one. Therefore, we design the multi-view module (MV) to be utilized on CL-I2C, which takes into account the multiple interests of the user. In addition to that, the user's interests in different spaces are multi-granular. Thus, we introduce the multi-granularity module (MG) on CL-I2C, which focuses on the user's both fine- and coarse-grained interests. To study the effects of multi-granularity and multi-view modules, we conduct extensive experiments on the datasets. According to Table \ref{tab:ab2}, we can see that both MV and MG have a positive effect on the model. And if they are combined together, the model performance will be improved even more.
\subsection{Case Study}
\begin{figure}[t]
    \centering\includegraphics[scale=0.53]{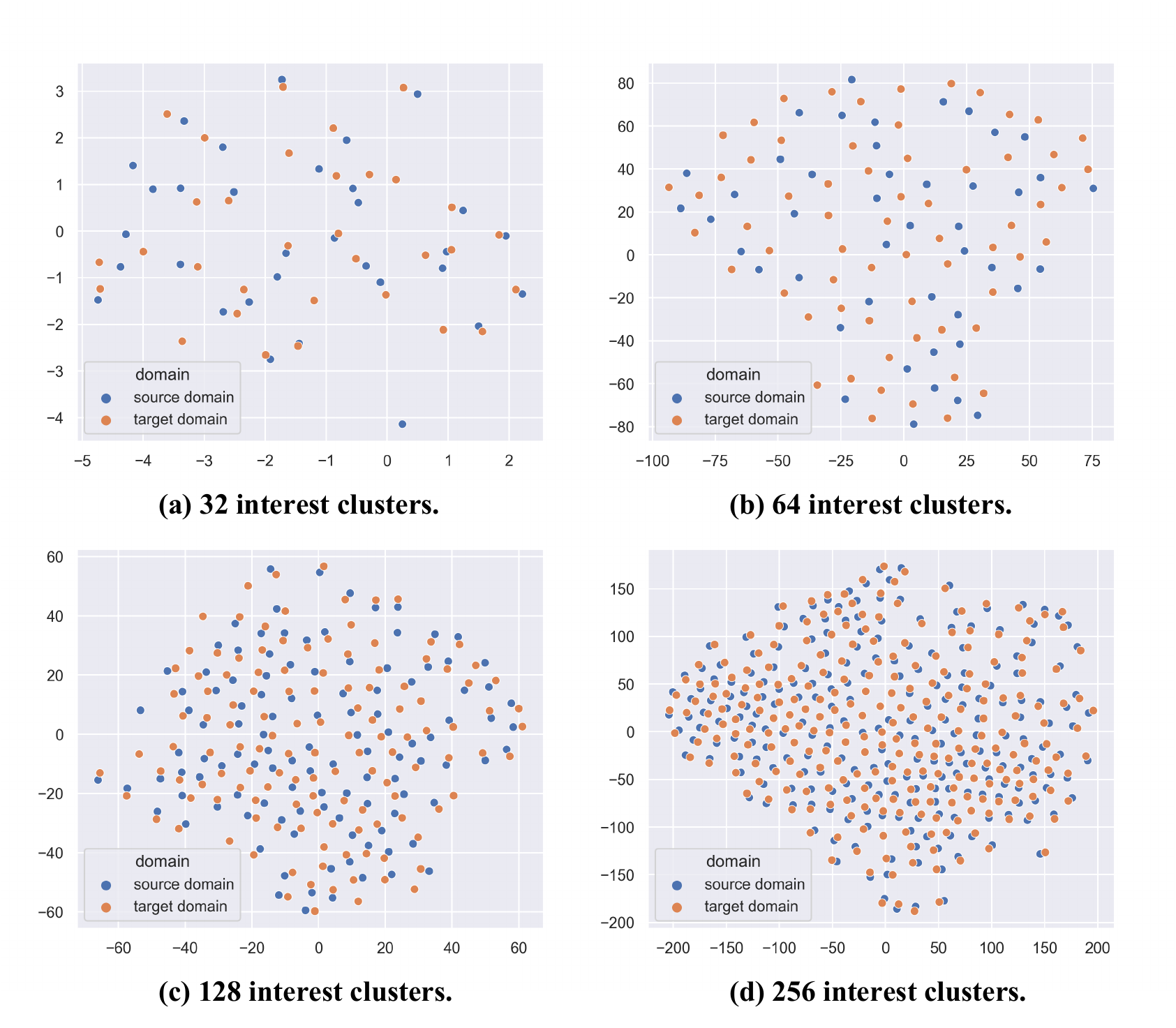}
    \caption{Visualization of the interest clusters in the source domain and the target domain. For (a), (b), (c) and (d), the numbers of clusters in both domains are the same, 32, 64, 128 and 256, respectively.}
    \label{fig:case}
\end{figure}
We study whether SITN can learn and transfer the interest invariance between domains. We initialize four interest spaces with different granularity and conduct experiments on the industrial dataset. These four interest spaces have the same numbers of clusters in the source domain and the target domain, 32, 64, 128 and 256, respectively. Then, we visualize the interest clusters in the both domains via t-SNE \cite{van2008visualizing}. As shown in Figure \ref{fig:case}, we can find that the interest clusters are evenly distributed, and almost each interest cluster in the source domain has an interest cluster in the target domain that is very close to it. Moreover, the distributions of the interest clusters in the two domains are also relatively close. It reflects the fact that SITN does learn the invariant knowledge of interests between domains, and thus improves the model performance.
% \subsection{Online Experiments}
% We conduct a careful online A/B testing on a real-world micro-video recommendation platform. The goal of the recommendation platform is to increase users' stickiness and activity. The experiment duration is three days, from 2022-0802 to 2022-0804, and is under the bucket tests. One bucket is selected for baseline and another bucket for our model. Each bucket serves about 0.2 million users per day. And the current baseline for the online model is BST \cite{chen2019behavior}. During three days of A/B testing, compare with the baseline, our method contributes up to 10.84\% number of browsing videos per user and 2.96\% dwell time per user. These results indicate that our method is more efficient than the baseline in terms of increasing the users' stickiness and activity and further demonstrate its value in real-world scenarios.

\section{Conclusion}
In this paper, we identify the significance of the interest invariance between domains in cross-domain recommendation. We propose a novel method called Self-supervised Interest Transfer Network (SITN), which employs instance-to-instance and instance-to-cluster contrastive learning simultaneously. Moreover, SITN also considers the user's multi-granularity and multi-view interests. With this paradigm, SITN can efficiently learn and transfer the interest invariance between domains. Extensive offline experiments on a large-scale industrial dataset and a public dataset indicate the effectiveness of the proposed method. 
% Additionally, we have deploy SITN on a real-world micro-video recommendation platform. 
And the online A/B testing results further demonstrate its practical value.

\section{Acknowledgments}
We thank the anonymous reviewers for their constructive comments to this paper. This work is supported by the Pioneer and Leading Goose R\&D Program of Zhejiang (2022C01243), and the Fundamental Research Funds for the Central Universities (Zhejiang University NGICS Platform).

\bibliography{aaai23.bib}
\end{document}